\documentclass[english,a4paper,eqno,11pt]{article}

\usepackage[english]{babel}
\usepackage{graphicx,epsfig,epic} 
\usepackage{pifont}      

\usepackage{xypic}

\usepackage{pict2e}

\font\tenbb=msbm10 at 12pt

\def\rR{\hbox{\tenbb R}}

\def\cal{\mathcal}

\def\gesp{\vskip1cm}
\def\esp{\vskip .6cm}
\def\pesp{\vskip .3cm}

\def\di{\displaystyle}

\newtheorem{thm}{Theorem}
\newtheorem{defn}{Definition}

\newtheorem{prop}{Proposition}

\newtheorem{cor}{Corollary}
\newtheorem{Exp}{Example}
\newtheorem{rem}{Remark}

\begin{document}

\title {Infinity of Geodesics in an Homogeneous and Isotropic\\ Expanding Space-Time}

\author{\small Fay\c cal BEN ADDA{\footnote{New York Institute of Technology.
Email: f\_benaddafr@yahoo.fr}}\quad H\'el\`ene PORCHON{\footnote{UFR 929 Math\'ematiques, Universit\' e Pierre et Marie Curie. Email: helene.porchon@etu.upmc.fr}}}

\maketitle


\esp

{\small Abstract. In this paper we construct a discrete simulation of an expanding homogeneous and isotropic space-time that expands via expansion of its basic elements to figure out properties and characteristics of such a space-time and derive conclusions. We prove that in such an expanding space-time, the geodesics are curved and more precisely, they fluctuate on the boundaries of the expanding basic elements. The non existence of privileged expansion direction leads to the existence of an infinity of fluctuating geodesics between any two locations in this space-time, that provides a prediction of polarization in geometric optics, and a prediction of an earlier acceleration of the expansion as for the cosmic inflation model. This simulation is a case study and an example of space-time with variable topology using the principle of variation of topology via a transformation that creates holes.}


\esp

\section{Introduction}
Everything in our world is perceived through the intermediary of light despite its enigmatic nature. All our observations come from the interaction and reflection of light with matter. To provide a comprehensive understanding of the universe expansion and elucidate matter interaction and distribution with space-time expansion, it is helpful to model the observed expansion of the space-time.\pesp

From observations we know that "{\it all galaxies are receding from each other with a velocity proportional to their distance}": this was the major valuable experimental information brought to our knowledge by E.P. Hubble in 1925 \cite{Chap3HUB1},\cite{Chap3HUB2},\cite{Chap3HUB3}, which was incomprehensible since it provided a big contradiction with the interpretation of the recession velocity of faraway galaxies that could exceed the velocity of light. In 1927, G. Lema\^{\i}tre \cite{Chap3GL} was the first who related the movement of galaxies to the space-time expansion: the whole space is stretched in all directions, which creates the space-time expansion, and then induces the recession movement of galaxies all around. To understand the free motion in that expanding space-time one needs to determine the geodesics of light on it since it is known that a freely moving physical system follows the geodesics of the space that optimize its extremum.\pesp

Light follows the geodesics of the space that represent the shortest distance between two different points, and to observe how it crosses that variable geometry within a concrete model will provide the print of the geometry of the expanding space-time. The main difficulty resides in the nature of the expansion of the space-time. How can it be modeled locally to provide a clear global picture that conveys observations (homogeneity and isotropy), especially when the variation of the space-time curvature is not only from one point to another but within the same point? Do we need initial conditions to start with? The metric that expresses the line element between two nearby events at a given point of the space-time must be variable to express the global expansion of the space via its local expansion. \pesp

\subsection{Main Objective of the Simulation}

The main objective of this work is:

i) to provide a discrete simulation of an expanding homogeneous and isotropic space-time that expands via expansion of its basic elements to figure out properties and characteristics of such an expanding space-time and derive conclusions that could be an asset for a better comprehension of the expansion of our universe, which leads, based on physical simulation, to obtain interpretation of the phenomena. Simulating an expanding space-time does not mean modeling our proper universe, or pretending that this is the model of our universe. This simulation is rather a case study to understand the initial condition and mechanism of the expansion, and leads to reproduce an expanding space-time that conveys the principles of cosmology such as homogeneity and isotropy. Other simulations might be developed with other conditions and mechanism that remain consistent with observations. Nevertheless, this simulation incorporates the principles of cosmology in a discrete expansion of a continuous space-time to observe what happens to the geodesics of the space-time as it expands via simultaneous expansion of its basic elements. The cause of this expansion, or why the basic elements expand simultaneously is out of the scope of this work. The initial condition and mechanism are chosen to determine the consequences on geodesics and then on matter distribution. Predictions and quantitative comparisons with experience will be discussed.

ii) to construct an example of space-time with variable topology using the principle of variation of topology via transformation that creates holes, which will constitute an application of variable topology that generates space expansion.

\subsection{Initial Condition of the Simulation and Mechanism}

An Euclidian space-time with zero curvature, set of accumulated points without depth (the limit entity is called basic element) is taken as a starting stage of the modeled space-time. As the space-time expands via the expansion of its basic elements, the Euclidian space-time will expand via simultaneous expansion of its basic elements, that is to say points increase simultaneously their depth (see Remark \ref{R001}) in all directions as balls. An infinite family of packed expanding balls is obtained that models the observed expansion in all directions. Taking into account the density of the space-time, other new basic elements with different sizes will expand in any interstice, which creates an Apollonian gasket of packed expanding balls with different sizes. The appearance of Apollonian gasket between any three packed basic elements creates the fractal character of the space-time and simulates locally the non homogeneity of the expanding space-time; meanwhile the simultaneous expansion of the primordial basic elements simulates the homogeneity and isotropy of the expanding space-time at large scale which is in concordance with the cosmological principles. In this kind of variable deformation of the space-time, due to the simultaneous expansion of its basic elements, it is not simple to set up equations that govern the geodesics on it, we have to observe and isolate fundamental characteristics from the modeled expanding space-time. \pesp

\begin{rem}\label{R001}
A simulation is known as the imitation of the operation of a physical system over time (\cite{Chap3BCNN},\cite{Chap3SB}). The model in this work represents the space-time expansion, and the simulation represents the mechanism (the operation) of the space-time expansion over time to identify the eventual characteristics and  effects of alternative conditions and course of transformation. The mechanism of expansion of balls can be understood by increasing the radius of the balls. However, how can one understand the expansion of a dimensionless point (without depth) to a ball? The expansion of points by increasing their depth in all directions as balls is an illustration of the transformation of a basic element of the Euclidian space-time to the basic element of the space-time given by an infinite family of packed balls. This transformation can be explained using two approaches:

(i) A point is a limit entity without depth, and a ball with radius $r$ is the accumulation of an infinity of points. The process that makes one point transformed into an infinity of accumulated points to form a ball with radius $r$ is possible by an infinite replication of points and accumulations. The process of replacing simultaneously each point in the Euclidian space-time by an infinity of accumulated points that form a ball with radius $r$ creates the expansion of the space-time. It is an expansion with self-similarity: the space-time is the accumulation of an infinity of points, and a point becomes an accumulation of another infinity of points (a 3 dimensional open ball is homeomorphic to $\rR^3$).

(ii) A point can be considered as a ball of dimension 0, then the process that makes a ball of dimension 0 transformed into a ball of dimension 3 can be understood by a dimensional expansion of balls.
\end{rem}

Many investigations that concern packing of spheres on a space with constant curvature can be found in \cite{Chap3BL},\cite{Chap3BOR},\cite{Chap3BUR},\cite{Chap3GA},\cite{Chap3LA},\cite{Chap3LI},\cite{Chap3MIL},\cite{Chap3MO},\cite{Chap3RA},\cite{Chap3RO},\cite{Chap3SC},\cite{Chap3SO}, without mentioning the references from physics and chemistry. However a geometrical definition of the space-time with variable curvature that conveys an expanding space-time via the expansion of its basic elements can be found in the fractal manifold model \cite{Chap3BF1},\cite{Chap3BF2}.\pesp

\subsection{Plan}
in this work we prove that in an expanding space-time the geodesics are curved by the considered expansion and more precisely they fluctuate on the boundaries of the expanding basic elements, which induces a variable curvature of the space, meanwhile the appearance of  Apollonian gasket induces a variable topology. These variations model the distribution of matter at large scale and make the geometry invisible.\pesp

The plan of this paper is as follow: in section 2, we define a discrete quantification that leads to an expanding space-time via the expansion of its basic elements, and we define a metric that incorporates this quantification. In section 3, we define the notion of contact between basic elements and we study their properties. We define the continuous curved paths that pass via the expanding basic elements, we study the geodesics and we give an example of fluctuating geodesics. We conclude this section by proving the existence of an infinity of fluctuating geodesics between any two locations. In section 4, we discuss the invisibility of the geometry, the variable curvature and its effect on matter distribution at large scale. In section 5, we presents some applications in physics.

\gesp


\section{Expanding Space-Time and Metric}

\subsection{Expansion Quantification}

Let us consider an Euclidian space-time (the primordial space) as a starting stage of the modeled expanding space-time (set of accumulated dimensionless basic elements without depth). Let us suppose that this primordial space expands via the expansion of its basic elements by increasing their depth simultaneously in all directions as open balls. We denote this expanding space-time by $\cal M$. To simulate the evolution of the expanding space-time $\cal M$ we define a discrete expansion of $\cal M$ via simultaneous expansion of its basic elements by using a quantified expansion indexed by $n\geq0$, where the primordial space is obtained for $n=0$ denoted by $step(0)$, and as $n$ increases the space-time $\cal M$ expands step by step to reach the present space-time at the $step(n)$. The step expansion of the basic elements of the space-time $\cal M$ is defined as follow:

\begin{defn}\label{Chap2S}
Let $P_0$ be a primordial basic element of the expanding space-time $\cal M$ at the $step(0)$  (a dimensionless basic elements without depth), and $P_{n}$ be its representation in  $\cal M$  at the $step(n)$ for all $n>0$. The expansion of the basic element $P_n$ verifies:

i) for all $n>0$,\quad  $P_{n}$ is an expanding open ball of radius $r_n>0$.

ii)  for all $n\geq0$,\quad $P_{n}\subset P_{n+1}$.

iii) for all $n>0$, any non empty subset  $S_n$ of $P_n$ expands symmetrically in all directions into a subset $S_{n+1}$ of $P_{n+1}$ such that $S_n\subset S_{n+1}$  .

\end{defn}

The increase of the distance between any two distant basic elements of the discrete expansion of the space time ${\cal M}$ is quantified from the $step(1)$ to the $step(n)$ as follow:

\begin{defn}\label{S0}
Let $P_1$ and $Q_1$ be two distant basic elements of the expanding space-time $\cal M$ at the $step(1)$ with distance $L_1>0$. Let $P_{n}$ and $Q_{n}$ be their respective representation at the $step(n)$ for all $n>1$. The distance between  $P_n$ and $Q_n$ at the $step(n)$ is defined by $L_n$ for all $n>1$ such that
\begin{equation}
    L_{n} = a_{n}\ L_{n-1}, \quad \forall n >1
\end{equation}
where the sequence $(a_i)_{i\geq1}$  satisfies the following conditions:

i) for all  $i>1$,\quad $a_i>1$.

ii) for all $i>1$,\quad $a_{i+1}<a_i.$

iii) the product $\di\prod_{i=1}^{n} a_{i+1}$ is convergent. For $a_1=1$ we have:
\begin{equation}\label{C3PROD}
\di L_{n}=L_1\Big(\prod_{i=1}^{n} a_i\Big),\quad \forall n\geq1,
\end{equation}
we call $a_i$ the $i^{th}$ step-expanding parameter and $ \prod_{i=1}^{n} a_i$ the expanding parameter at the $step(n)$.
\end{defn}

\begin{rem}\label{Rem2}

i) The previous quantification starts from the $step(1)$ to conserve the convergence of equality (\ref{C3PROD}): indeed, let $P_0$ and $Q_0$ be two distant primordial basic elements of the expanding space-time $\cal M$ at the $step(0)$ with finite distance $L_0>0$, and let $P_1$ and $Q_1$ be their transformation after expansion from the $step(0)$ to the $step(1)$, where  $L_1=d(P_1,Q_1)$ is their distance after transformation. Since there exists an infinity of points between the primordial points $P_0$ and $Q_0$, then after simultaneous transformation of the points $P_0$ and $Q_0$ and all the intermediary points, the points $P_1$ and $Q_1$ at the $Step(1)$ will be separated with an infinity of aligned open balls of dimension 3, which makes $L_1=d(P_1,Q_1)$ infinite.

ii) Any finite distance between $P_0$ and $Q_0$ at the $Step(0)$ becomes infinite after transformation at the $Step(1)$ and any finite distance between $P_n$ and $Q_n$ at the $Step(n)$ remains finite after transformation at the $Step(n+1)$ for all $n>0$.

iii) The main objective of this simulation is to identify the characteristics and  effects of the local transformation of the basic element on matter. The divergence of the distance at the first step does not affect the main objective, however it can provide an interpretation of the existence of acceleration at the first step as the model of inflation in cosmology (this will be discussed in section 5).

iv) If we consider the sequence $(a_i)_{i\geq1}$ of parameters as new dimensions, it can be seen from the metric (\ref{M1}) that these new dimensions are not perpendicular as the $x$, $y$ and $z$ dimensions. The above quantification from the $step(1)$ to the $step(n)$ describes a discrete unitary increase of dimensions, depending on the new parameters, meanwhile from $step(0)$ to $step(1)$, the number of dimensions increases from 0 to 3 (the perpendicular dimensions).
\end{rem}

\begin{Exp}It is not difficult to see that the following sequence satisfies the conditions of the previous definition:
\begin{equation}
a_i=\left\{
         \begin{array}{ll}
           1+\di\frac{1}{(i-1)^2} & \hbox{for}\quad i>1 \\
           1 & \hbox{for}\quad i=1.
         \end{array}
       \right.
\end{equation}
\end{Exp}

\subsection{The Space-Time Metric}

In this framework there is no concrete use of the metric, nevertheless we will give the metric that defines the distance between two space-time events in the expanding space-time $\cal M$ for better understanding of the formulation of the space-time expansion via discrete expansion of its basic elements from one step to another. The metric of the expanding space-time $\cal M$ is given at the $step(n)$ by
\begin{equation}\label{M3}
d\sigma^2_n=g_{ij}dy^idy^j\qquad i,j=1,2,3,4
\end{equation}
where the $y^i$ for $i=1,2,3,4$, represent the curvilinear coordinates and where  $g_{ij}$ for  $i,j=1,2,3,4$, are symmetric functions of $y^i$. If the coordinates are reduced to rectangular Cartesian coordinates, the only non zero coefficients of the metric are
\begin{equation}
g_{11}=g_{22}=g_{33}=-\prod_{i=1}^n a_i, \qquad g_{44}=+1.
\end{equation}

In the expanding space-time $\cal M$, the proper distance along a curve is measured for $j=1,2,3$  by $\Big ( \prod_{i=1}^na_i\Big)x_j$ rather than $x_j$ at the $step(n)$, then in rectilinear coordinates the metric of $\cal M$ has a reduced expression at the $step(n)$ given by

\begin{equation}\label{M1}
d\sigma^2_n=c^2dt^2-\Big ( \prod_{i=1}^na_i\Big)^2(dx_1^2+dx_2^2+dx_3^2)\qquad \quad \forall n\geq 1.
\end{equation}

\begin{rem}\label{Rem1}

In a complete universe and for all references, the geodesics are locally curves that minimize length, more precisely the geodesics are the extremum of the optimization
\begin{equation}\label{G1}
\delta\int _{z_0}^{z_1}d\sigma=0,
\end{equation}
where $d\sigma^2$ is the general metric of the considered space-time (line element). In a classical approach a freely moving body follows a geodesic of space-time that optimizes the extremum of (\ref{G1}) (which corresponds to the initial condition of movement). The classical approach to find the geodesics is only based on the use of the local metric. Nevertheless in this framework the geodesic of a freely moving body in the expanding space-time $\cal M$ will be studied and found using Definition \ref{Chap2S} and Definition \ref{S0} and (\ref{G1}).

\end{rem}

\section{Expanding Space-Time and Geodesics}

Discovering matter distribution in an expanding space-time amounts to determine the geodesics of light in such an expanding space-time.

\subsection{Contacts and Basic Elements in an Expanding Space-time}

Since the expanding space-time $\cal M$ expands via discrete expansion of its basic elements in all directions, and since these basic elements expand as packed balls, then we need to study contacts between different packed open balls, boundaries, and continuous paths on spheres. For this study we use a classical distance $d$ to determine the distance between points, or the distance between points and sets of points on those open expanding balls. We introduce the following:

\begin{defn}\label{D6}
Let $P_n$ and $Q_n$ be two basic elements of the expanding space-time $\cal M$ at the $step(n)$ for all $n> 0$.
The basic elements $P_n$ and $Q_n$ have a simple contact at the point $s_n$ at the $step(n)$ for all $n> 0$
if:

i) $\forall n> 0$,\quad $P_n\cap Q_n=\emptyset$.

ii) $\forall n>0$,\quad   $d(s_n,P_n)=0$ \quad and \quad$d(s_n,Q_n)=0$.
\end{defn}

\begin{prop}
If $P_n$ and $Q_n$ are two basic elements of the expanding space-time $\cal M$ with simple contact at the $step (n)$ for all $n> 0$, then for all $n> 0$,\quad $d(P_n,Q_n)=0$.
\end{prop}

{\it Proof:} Let $P_n$ and $Q_n$ be two basic elements of the expanding space-time $\cal M$ with simple contact $s_n$ at the $step (n)$ for all $n> 0$. For all $n> 0$, for all $x\in P_n$ and $y\in Q_n$ we have
$$d(x,y)\leq d(x, s_n)+ d(s_n, y),$$
then
$$\inf_{y\in Q_n}d(x,y)\leq \inf_{y\in Q_n}d(x, s_n)+ \inf_{y\in Q_n}d(s_n, y),$$
which gives
$$\inf_{y\in Q_n}d(x,y)\leq d(x, s_n)+ d(s_n, Q_n),$$
moreover
$$\inf_{x\in P_n}\inf_{y\in Q_n}d(x,y)\leq \inf_{x\in P_n}d(x, s_n)+ \inf_{x\in P_n}d(s_n, Q_n),$$
then
$$\inf_{x\in P_n}\inf_{y\in Q_n}d(x,y)\leq d(P_n, s_n)+ d(s_n, Q_n).$$

Since $s_n$ is the simple contact of the  basic elements $P_n$ and $Q_n$, then
$$\inf_{x\in P_n}\inf_{y\in Q_n}d(x,y)\leq 0$$
which gives
$$\inf_{x\in P_n}\inf_{y\in Q_n}d(x,y)=0,$$
that is to say $d(P_n, Q_n)=0$ for all $n> 0$.

\begin{prop}
If $P_n$ and $Q_n$ are two basic elements of the expanding space-time $\cal M$ with simple contact at the $step(n)$ for all $n> 0$, then for all $n> 0$ this simple contact is unique.
\end{prop}

{\it Proof:}
If $P_n$ and $Q_n$ are two basic elements of the expanding space-time $\cal M$ with a simple contact $s_n$ at the $step(n)$ for all $n> 0$, then $d(s_n,P_n)=0$ and $d(s_n,Q_n)=0$ with $P_n\cap Q_n=\emptyset$. Suppose that $t_n$ is another simple contact between $P_n$ and $Q_n$, then by the same we have $d(t_n,P_n)=0$ and $d(t_n,Q_n)=0$ with $P_n\cap Q_n=\emptyset$. Since $d(s_n,P_n)=d(s_n,\overline{P}_n)$ and $d(s_n,Q_n)=d(s_n,\overline{Q}_n)$
then
$$
\left\{
  \begin{array}{ll}
    d(s_n,P_n)=0, & \Leftrightarrow\quad s_n\in \overline{P}_n\\
    d(s_n,Q_n)=0, & \Leftrightarrow\quad s_n\in \overline{Q}_n
  \end{array}
\right.
$$
which gives $s_n\in \overline{P}_n\cap \overline{Q}_n$. Since $P_n\cap Q_n=\emptyset$\quad then\quad $s_n\in \partial\overline{P}_n\cap \partial\overline{Q}_n$.
The same steps for the simple contact $t_n$ gives that $t_n\in \partial\overline{P}_n\cap \partial\overline{Q}_n$. Since $\partial\overline{P}_n=\overline{P}_n\backslash P_n$ and $\partial\overline{Q}_n=\overline{Q}_n\backslash Q_n$ are two spheres with same radius, such that $\partial\overline{P}_n\cap \partial\overline{Q}_n\neq\emptyset$ and $P_n\cap Q_n=\emptyset$, then $s_n=t_n$, which gives the uniqueness of the simple contact for all $n>0$.

\begin{defn}\label{D5}
A continuous curved path ${\cal C}_n$ has a continuous contact with the basic element $P_n$ at the $step(n)$ for all $n>0$ if:

i) $P_n\cap {\cal C}_n=\emptyset$, \quad$\forall n> 0$

ii) For all\quad $n>0$\quad there exists a closed connected subset ${\cal I}_n\subset {\cal C}_n$, such that for all $x \in{\cal I}_n$,\quad $d(x,P_n)=0$.
\end{defn}

\begin{defn}\label{Go}
A continuous curved path ${\cal C}_n$ has a geodesic continuous contact with the basic element $P_n$ at the $step(n)$ for all $n>0$ if
${\cal C}_n$ has a continuous contact with the basic element $P_n$ for all $n>0$, where the closed connected subset ${\cal I}_n\subset {\cal C}_n$ such that for all $x \in{\cal I}_n$\quad $d(x,P_n)=0$, is a geodesic arc on the boundary of $P_n$ .

\end{defn}

\begin{defn}\label{Via}
Let $P_n$ be a basic element of the expanding space-time $\cal M$ at the $step(n)$ for all $n> 0$. A continuous curved path ${\cal C}_n$ passes via the basic elements $P_n$ at the $step (n)$ for all $n>0$ if ${\cal C}_n$ satisfies one of the following conditions:

i) $\forall n> 0$,\quad${\cal C}_n\cap P_n\not=\emptyset$.

ii) $\forall n> 0$,\quad${\cal C}_n$ has a continuous contact with $P_n$.
\end{defn}

\begin{defn}
Let $P_n$ and $Q_n$ be two basic elements of the expanding space-time $\cal M$ with a simple contact at the $step(n)$ for all $n> 0$. A continuous curved path ${\cal C}_n$ passes via the basic elements $P_n$ and $Q_n$  with continuous contacts at the $step(n)$ for all $n>0$ if the path ${\cal C}_n$
has a continuous contact ${\cal I}_n$ with $P_n$ and a continuous contact ${\cal J}_n$ with $Q_n$ such that ${\cal I}_n\cap{\cal J}_n\not=\emptyset$.
\end{defn}

\begin{prop}\label{Pr3}
Let $P_n$ and $Q_n$ be  two basic elements of the expanding space-time $\cal M$ with a simple contact $s_n$ at the $step(n)$ for all $n> 0$.
If $\ {\cal C}_n$ is a continuous curved path that passes via the basic elements $P_n$ and $Q_n$ with continuous contact ${\cal I}_n$ and ${\cal J}_n$ respectively for all $n>0$,
then \quad  ${\cal I}_n\cap {\cal J}_n=\{s_n \}$ for all $n>0$.
\end{prop}

{\it Proof:} Let $P_n$ and $Q_n$ be  two basic elements of the expanding space-time $\cal M$ with a simple contact $s_n$ at the $step(n)$ for all $n> 0$. If $\ {\cal C}_n$ is a continuous curved path that passes via the basic elements $P_n$ and $Q_n$ with continuous contact ${\cal I}_n$ and ${\cal J}_n$ respectively for all $n>0$, then
$$
\left\{
  \begin{array}{ll}
   {\cal C}_n\cap\partial\overline{P}_n={\cal I}_n , & \forall n>0\\
    {\cal C}_n\cap\partial\overline{Q}_n={\cal J}_n, & \forall n>0.
  \end{array}
\right.
$$

The two basic elements $P_n$ and $Q_n$ have a simple contact $s_n$ at the $step(n)$ for all $n> 0$, and verify $\partial\overline{P}_n\cap \partial\overline{Q}_n\neq\emptyset$, with $P_n\cap Q_n=\emptyset$, then

${\cal I}_n\cap{\cal J}_n=\Big({\cal C}_n\cap\partial\overline{P}_n\Big)\cap\Big({\cal C}_n\cap\partial\overline{Q}_n\Big)={\cal C}_n\cap\Big(\partial\overline{P}_n\cap\partial\overline{Q}_n\Big)={\cal C}_n\cap\{s_n\}=\{s_n\}$, since the continuous curved path $\ {\cal C}_n$ passes via the basic elements $P_n$ and $Q_n$ with continuous contact.

\begin{prop}\label{Pr2}
Let $Q_n$ be a basic element of the expanding space-time $\cal M$ at the $step(n)$ for all $n>0$, and\quad ${\cal C}$\quad a continuous path.
If $Q_n\cap {\cal C}\not=\emptyset$, then for all $n\geq 0$, $Q_n\cap {\cal C}$ is an expanding subset of $Q_n$.
\end{prop}

{\it Proof:} If $Q_n\cap {\cal C}\not=\emptyset$ then there exists a non empty set $S_n$ such that $S_n\subset Q_n$ and $S_n\subset {\cal C}$.
If $Q_n$ is expanding into $Q_{n+1}$ , then $S_n$ is expanding into $S_{n+1}\subset Q_{n+1}$ (by Definition \ref{Chap2S}) such that $S_n\subset S_{n+1}$, which concludes the proof.

\subsection{Geodesics of Light in an Expanding Space-Time}

More precisely let us consider a geodesic of light that passes from a source $S$ to a point $B$ in the expanding space-time $\cal M$. The light geodesic in the primordial space-time at the $step(0)$ was a straight line between $S$ and $B$ (Remark \ref{Rem1}).  Since the space-time is expanding through the simultaneous expansion of its basic elements, any point (dimensionless basic element without depth) of the space located between the source $S$ and the end point $B$ expands.

Indeed, as the space-time expands through the expansion of its basic elements, all basic elements that constitute the light ray path of the line segment $[SB]$ will expand as aligned balls with simple contacts.


Assuming that the geodesics are one dimension continuous paths as the space-time expands, we have the following results:

\begin{thm}\label{Chap2L2}
The light geodesics between two distinct locations $S$ and $B$ in the expanding space-time $\cal M$ verifies the following:

i) they are given by the straight line segment $[SB]$ in the primordial space-time at the $step(0)$.

ii) they pass via a family of aligned expanding basic elements, located between $S$ and $B$, with geodesic continuous contacts at the $step(n)$ for all $n>0$.

\end{thm}

{\it Proof:} i) For $n=0$, the expanding space-time $\cal M$ is flat with basic elements given by points, dimensionless dots without depth, and the geodesics are defined by straight lines (Remark \ref{Rem1}), that is to say that the light geodesic between $S$ and $B$ is given by the straight line segment $[SB]$ at the $step(0)$.

ii) For $n>0$, let us consider all the basic elements of the expanding space-time $\cal M$ that constitute the light ray path of the line segment $[SB]$ at the $step (0)$. As $n$ increases, those basic elements expand as aligned balls with simple contact. To find the geodesic location of light between two distinct locations $S$ and $B$, is sufficient to find the geodesic location via two expanding basic elements with simple contact at the $step(n)$ for all $n>0$.

\begin{figure}[h]
\begin{center}
\includegraphics[width=7cm]{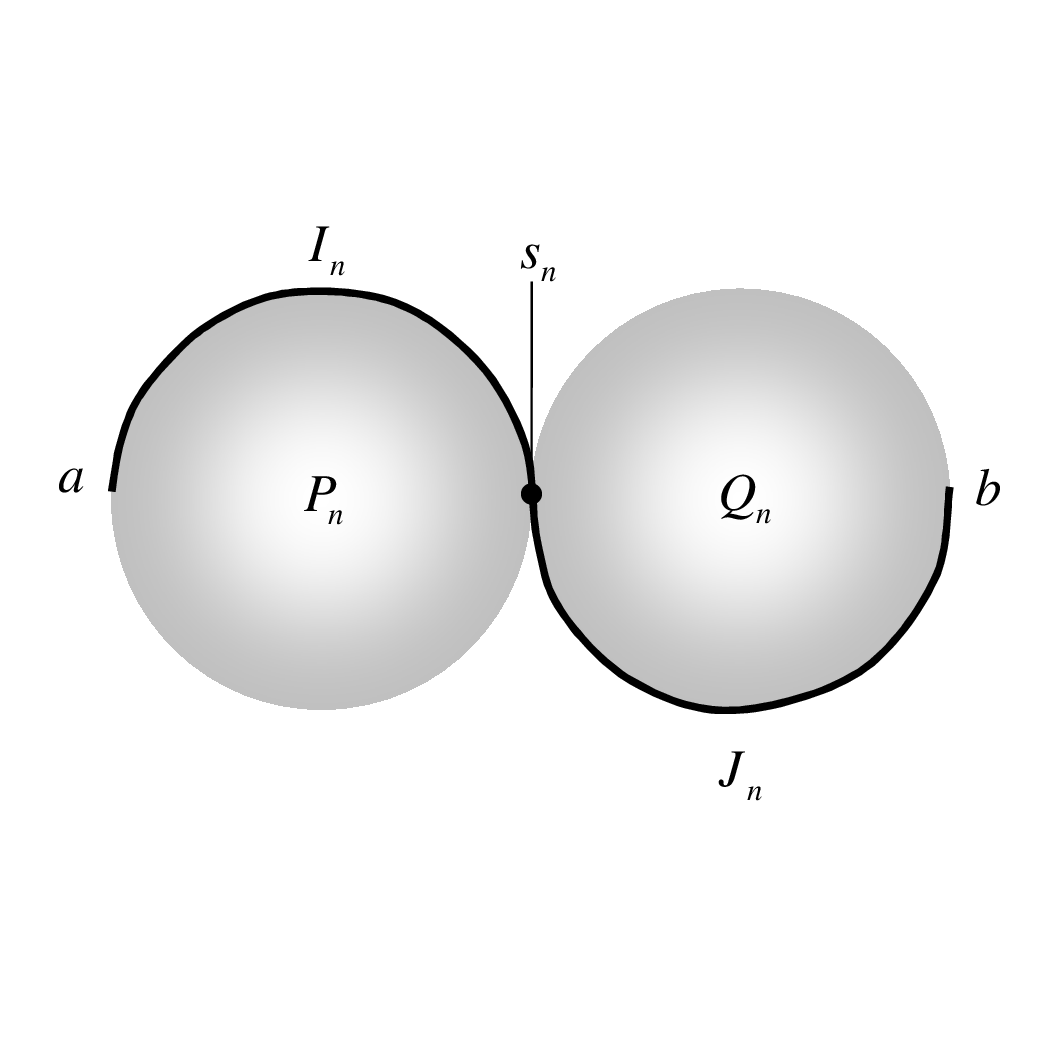}
\centering\caption{ {\footnotesize  Illustration of a geodesic passing via two expanding basic elements $P_n$ and $Q_n$ at the $step(n)$.}}\label{Light2}
\end{center}
\end{figure}

Let us consider a continuous curve $\gamma_n$, which represents the geodesic of light that passes from the location $S$ to the location $B$ via two basic elements $P_n$ and $Q_n$ at the $step(n)$ for all $n>0$, and let us suppose that the basic elements $P_n$ and $Q_n$ have a simple contact $s_n$ at the $step(n)$ for all $n>0$. Let $a$ and $b$ be two antipodal points to $s_n$ with respect to the boundary of $P_n$ and $Q_n$ respectively (see Fig.\ref{Light2}). If the geodesic $\gamma_n$ passes throughout $P_n$ and $Q_n$ using the points $a$, $s_n$, and $b$, then

\begin{equation}
\gamma_n\cap P_n\not=\emptyset\qquad\hbox{and}\qquad\gamma_n\cap Q_n\not=\emptyset.
\end{equation}

Therefore there must exist two subsets ${\cal C}_1$ and ${\cal C}_2$ of $\gamma_n$ such that ${\cal C}_1=\gamma_n\cap P_n$ and ${\cal C}_2=\gamma_n\cap Q_n$. Since $P_n$ and $Q_n$ expand in all direction as well as any subset of  $P_n$ and $Q_n$, then by Proposition \ref{Pr2} the subset ${\cal C}_1$ expands together with $P_n$ (respectively ${\cal C}_2$  with $Q_n$), which yields that  $\gamma_n$ is not anymore a curve since the subsets ${\cal C}_1$ and ${\cal C}_2$ of $\gamma_n$ are expanding in all direction as the basic elements expand.

Therefore the light geodesic $\gamma_n$ must not cross any part of the basic elements $P_n$ and $Q_n$ for all $n>0$, which requires

\begin{equation}\label{I1}
\gamma_n\cap P_n=\emptyset\qquad\hbox{and}\quad\gamma_n\cap Q_n=\emptyset\qquad\forall n>0,
\end{equation}
then by Definition \ref{Via} the geodesic $\gamma_n$ passes via $P_n$ and $Q_n$ with continuous contacts at the $step(n)$.

Since the light geodesic $\gamma_n$ is the shortest continuous path that the light follows to pass via the basic elements $P_n$ and $Q_n$, through antipodal points $a$, $s_n$, and $b$, then there exists a geodesic arc between the antipodal points $a$ and $s_n$ on the closure $\overline{P}_n$ of the basic element $P_n$, as well as a geodesic arc between the antipodal points $s_n$ and $b$ on the closure $\overline{Q}_n$ of the basic element $Q_n$, such that the geodesic $\gamma_n$ is a geodesic continuous contact.

That is to say for all $n>0$ there exist a closed connected subset ${\cal I}_n\subset\gamma_n$ given by the geodesic on $\overline{P}_n$ (the smallest sphere that contains $P_n$) between the antipodal points $a$ and $s_n$, and a closed connected subset ${{\cal J}_n\subset\gamma_n}$ given by the geodesic on $\overline{Q}_n$ (the smallest sphere that contains $Q_n$) between the antipodal points $s_n$ and $b$ at the $step(n)$, such that
\begin{equation}\label{I2}
\forall x\in{\cal I}_n,\quad d(x,P_n)=0\qquad\hbox{and}\qquad\forall x\in{\cal J}_n,\quad d(x,Q_n)=0.
\end{equation}
with
\begin{equation}
{\cal I}_n\cap{\cal J}_n=\{s_n\}.
\end{equation}
as illustrated in Fig.\ref{Light2}. Then $\gamma_n$ has a geodesic continuous contact with $P_n$ and $Q_n$ at the $step(n)$ for all $n>0$. This leads to the conclusion that between the two locations $S$ and $B$ the light follows a geodesic continuous contact with the family of all aligned expanding basic elements. That is to say the geodesic fluctuates and passes through simple contacts of aligned basic elements between the two locations $S$ and $B$, and its is not difficult to see that this geodesic tends to a straight line segment as the radius of the open balls tends to zero for $n=0$.
\begin{cor}\label{cor4}
The expansion of the basic elements of the expanding space-time $\cal M$ bends the geodesics of the light.
\end{cor}

{\it Proof:} As direct consequence of the Theorem \ref{Chap2L2}, since the light follows a geodesic $\gamma_n$ of the space-time at the $step(n)$ for all $n\geq0$, and that the space-time expands via the expansion of its basic elements, and since each geodesic continuous contact on each basic element (part of the geodesic $\gamma_n$) is varying together with the variation of the curvature of the expanding spheres that contain the basic elements, then the expansion of the basic elements bends the geodesics of light. They were straight lines at the primordial space-time (at $step(0)$) and become curved geodesics with local variable curvature (the curvature of an expanding sphere varies as the radius varies). \pesp

\begin{cor}\label{C1}
In the expanding space-time $\cal M$, the light geodesics fluctuate together with the space-time expansion, and are characterized by two directions:

i) a geodesic global direction that defines the light direction from initial position to end position.

ii) a geodesic local direction that fluctuates via boundaries of expanding basic elements at the $step(n)$ for all $n>0$.
\end{cor}

{\it Proof:} i) The proof is straight forward from i) of Theorem \ref{Chap2L2}.

ii) Using ii) of Theorem \ref{Chap2L2}, between any two basic elements $P_n$ and $Q_n$ with a simple contact $s_n$ for $n>0$, the geodesics are  continuous contacts that pass through the simple contact $s_n$ at the $step(n)$, therefore the geodesics will fluctuate via the boundaries of the basic elements $P_n$ and $Q_n$ and their simple contact $s_n$ for all $n>0$.\pesp

\begin{Exp} {\bf : Fluctuating geodesics}\pesp

The geodesics of light are given by piecewise differentiable paths that fluctuate via the boundaries of expanding basic elements (simulated by expanding balls). The differentiable parts of the geodesics are identified to be the geodesics on spheres given by arc of great circles. These parts of geodesics can be illustrated in 2 dimensions as follow:

Let us consider $N$ aligned balls ${\cal B}_{n,i}\Big(C_{n,i}, r_{n}\Big)$ for $i=0,\dots, N-1$ at the $step(n)$ of the expanding space-time $\cal M$ for all $n>0$, with centers given by $C_{n,i}=\Big((2i+1)r_{n}, 0\Big)$ for $i=0,\dots,N-1$ and radius given by $r_{n}=\Big ( \prod_{j=1}^{n}a_j\Big)r_1$, where $r_1>0$ is the radius of the balls at the $step (1)$ (just at the beginning of the expanding space-time expansion, $r_0=0$ at the $step(0)$ because the basic elements are points without depth and dimensionless).

The simple contacts of the $N$ aligned balls are given by the points $(2ir_{n},0)$ for $i=1\dots, N-1$.

\vskip1cm
\setlength{\unitlength}{1.cm}
\begin{picture}(5,5)

\put(0,2){\vector(1,0){14}}
\put(0,2){\vector(0,1){3}}
\put(-0.3,2){$0$}
\put(13.6, 2.2){x}
\put(0.2, 5){y}

\thinlines
\put(1,2){\circle{2}}
\put(3,2){\circle{2}}
\put(5,2){\circle{2}}
\put(7,2){\circle{2}}
\put(9,2){\circle{2}}
\put(11,2){\circle{2}}

\put(1,2){\circle*{.05}}
\put(3,2){\circle*{.05}}
\put(5,2){\circle*{.05}}
\put(7,2){\circle*{.05}}
\put(9,2){\circle*{.05}}
\put(11,2){\circle*{.05}}

\put(0.8,2.2){$r_n$}
\put(2.8,2.2){$3r_n$}
\put(4.8,2.2){$5r_n$}
\put(6.8,2.2){$7r_n$}
\put(8.8,2.2){$9r_n$}
\put(10.8,2.2){$11r_n$}

\put(0.8,0.5){${\cal B}_{n,1}$}
\put(2.8,0.5){${\cal B}_{n,2}$}
\put(4.8,0.5){${\cal B}_{n,3}$}
\put(6.8,0.5){${\cal B}_{n,4}$}
\put(8.8,0.5){${\cal B}_{n,5}$}
\put(10.8,0.5){${\cal B}_{n,6}$}

\linethickness{.7mm}
\put(1,2){\oval[1](2,2)[t]}
\put(3,2){\oval[1](2,2)[b]}
\put(5,2){\oval[1](2,2)[t]}
\put(7,2){\oval[1](2,2)[b]}
\put(9,2){\oval[1](2,2)[t]}
\put(11,2){\oval[1](2,2)[b]}

\put(2,-1){\shortstack[c]{\small{Figure 2:} \footnotesize \  Representation of a geodesic in 2D
for $N=6$ at the $step(n)$.}}

\end{picture}
\vskip 3cm

The geodesic in Fig.2 is the graph of the function $\varphi_{n}$ at the $step(n)$ defined by:
\begin{equation}\label{Geod}
\varphi_{n}(x)=\sum_{i=0}^{N-1}g_{n,i}(x)
\end{equation}
where
\begin{equation}
g_{n,i}(x)=\left\{
             \begin{array}{ll}
               \varphi_{n,i}(x) & \hbox{for}\quad x\in [2ir_n,2(i+1)r_n]\\
               0 & \hbox{for}\quad x\not\in [2ir_n,2(i+1)r_n]
             \end{array}
           \right.
\end{equation}
and
\begin{eqnarray}
  \nonumber \varphi_{n,i} :\ \rR&\longrightarrow& \rR \\
  x &\longmapsto & \varphi_{n,i}(x)\ =\ (-1)^i\ \sqrt{r_{n}^2 - \Big(x-(2i+1)r_{n}\Big)^2}
\end{eqnarray}
that verifies:

\begin{itemize}
  \item for $i=0,\dots, N-1$, the graph of $\varphi_{n,i}$ represents the geodesic between two antipodal points on the sphere of center
$C_{n,i}=\Big((2i+1)r_n, 0\Big)$ and radius $r_n$;
  \item for $i=0,\dots, N-1$,\  $\varphi_{n,i}$ is continuous on the closed interval \quad $[2ir_{n},2(i+1)r_{n}]$;
  \item for $i=0,\dots, N-1$,\  $\varphi_{n,i}$ is differentiable on the open interval \quad $]2ir_{n+1},2(i+1)r_{n}[$;
  \item for $i=0,\dots, N-1$,\  $\varphi_{n,i}$ is not differentiable at the points\quad $x_{n,i}=2ir_{n}$ and $x_{n,i+1}=2(i+1)r_{n}$.
\end{itemize}

\vskip 1.5cm
\setlength{\unitlength}{1.cm}
\begin{picture}(7,7)

\put(0,2){\vector(1,0){14}}
\put(14.5,2){$x$}
\put(0,6){\vector(1,0){14}}
\put(14.5,4){$x$}
\put(0,4){\vector(1,0){14}}
\put(14.5,6){$x$}
\put(0,2){\vector(0,1){6}}
\put(-0.3,2){$0$}
\put(0.2, 8){y}

\put(12.2, 2.2){\footnotesize$Step(n)$}
\put(11.6, 4.2){\footnotesize$Step(n-1)$}
\put(11.6, 6.2){\footnotesize$Step(n-2)$}


\linethickness{.7mm}
\put(1,2){\oval[1](2,2)[t]}
\put(3,2){\oval[1](2,2)[b]}
\put(5,2){\oval[1](2,2)[t]}
\put(7,2){\oval[1](2,2)[b]}
\put(9,2){\oval[1](2,2)[t]}
\put(11,2){\oval[1](2,2)[b]}

\put(1,2){\circle*{.05}}
\put(3,2){\circle*{.05}}
\put(5,2){\circle*{.05}}
\put(7,2){\circle*{.05}}
\put(9,2){\circle*{.05}}
\put(11,2){\circle*{.05}}

\put(0.8,2.2){$r_n$}
\put(2.8,2.2){$3r_n$}
\put(4.8,2.2){$5r_n$}
\put(6.8,2.2){$7r_n$}
\put(8.8,2.2){$9r_n$}
\put(10.8,2.2){$11r_n$}


\linethickness{.7mm}
\put(0.75,4){\oval[1](1.5,1.5)[t]}
\put(2.25,4){\oval[1](1.5,1.5)[b]}
\put(3.75,4){\oval[1](1.5,1.5)[t]}
\put(5.25,4){\oval[1](1.5,1.5)[b]}
\put(6.75,4){\oval[1](1.5,1.5)[t]}
\put(8.25,4){\oval[1](1.5,1.5)[b]}

\put(0.75,4){\circle*{.05}}
\put(2.25,4){\circle*{.05}}
\put(3.75,4){\circle*{.05}}
\put(5.25,4){\circle*{.05}}
\put(6.75,4){\circle*{.05}}
\put(8.25,4){\circle*{.05}}

\put(0.55,4.2){$r_{n-1}$}
\put(1.8,4.2){$3r_{n-1}$}
\put(3.3,4.2){$5r_{n-1}$}
\put(4.8,4.2){$7r_{n-1}$}
\put(6.3,4.2){$9r_{n-1}$}
\put(7.8,4.2){$11r_{n-1}$}


\linethickness{.7mm}
\put(.5,6){\oval[1](1,1)[t]}
\put(1.5,6){\oval[1](1,1)[b]}
\put(2.5,6){\oval[1](1,1)[t]}
\put(3.5,6){\oval[1](1,1)[b]}
\put(4.5,6){\oval[1](1,1)[t]}
\put(5.5,6){\oval[1](1,1)[b]}

\put(0.5,6){\circle*{.05}}
\put(1.5,6){\circle*{.05}}
\put(2.5,6){\circle*{.05}}
\put(3.5,6){\circle*{.05}}
\put(4.5,6){\circle*{.05}}
\put(5.5,6){\circle*{.05}}

\put(0.2,5.6){$r_{n-2}$}
\put(1.,6.3){$3r_{n-2}$}
\put(2.05,5.6){$5r_{n-2}$}
\put(3,6.3){$7r_{n-2}$}
\put(4.05,5.6){$9r_{n-2}$}
\put(5.05,6.3){$11r_{n-2}$}

\put(0,-1){\shortstack[l]{\small{Figure 3:} \footnotesize \ Fractal representation of the geodesic of light in 2D for $N=6$.
\ The geodesic  presents \\ \footnotesize different aspects at different steps of the space-time expansion (at the $step(n)$, $step(n-1)$ and \\
\footnotesize $step(n-2)$, the geodesic changes its form and length).\ \ The fractal character of the geodesic appears
  \\ \footnotesize together with the space-time expansion (as $n$ increases).}}
  \setcounter{figure}{3}
\end{picture}
\vskip1.5cm

As the basic elements expand in the expanding space-time (as $n$ increases), the radius $r_{n}$ increases, and then the points of non differentiability\quad $x_{n,i}=2ir_{n}$ of $\varphi_{n,i}$ change according to the $step(n)$ of the space-time expansion. The length $D_n=2r_{n}$ of the intervals of differentiability of $\varphi_{n,i}$ (which corresponds to the balls diameter at the $step(n)$) increases at each step of the space-time expansion.\pesp

At the $step(n)$ for $n>0$, the geodesics in the expanding space-time that represent the shortest path between initial point and end point is given by a finite sum of piecewise differentiable geodesics on spheres together with the existence of discrete points of non differentiability $\ x_{n,i}=2ir_{n}\ $ for $\ i=0,\dots, N$. Those points of non differentiability are  variable together with the space-time expansion.\pesp

Between the initial point $(0,0)$ and the end point $(2Nr_{n},0)$ there exists an infinity of geodesics in 3 dimensions (and $2^N$ geodesics in 2 dimensions). However all the geodesics have the same length given by $L_n=N\pi r_{n}$ that varies together with the space-time expansion (see Fig.3).\pesp
\end{Exp}

At the $step(0)$, the geodesics were straight lines of class ${\cal C}^1$, and as the space-time expands the geodesics lose their differentiability on a discrete set of points.
This example leads to the following theorem:

\begin{thm}\label{Theo}
If $\cal M$ is an expanding space-time that expands via  discrete expansion of its basic elements, then

i) the geodesics on $\cal M$ for all $n>0$ are sum of piecewise differentiable functions where the differentiable parts of the geodesics are great circle arcs;

ii) the geodesics on $\cal M$ admit a countable number of  points of non differentiability at the $step(n)$ for all $n>0$;

iii) for $n=0$ the geodesics on $\cal M$ do not admit any point of non differentiability.
\end{thm}

\pesp

{\it Proof:} To prove i) and ii) we will use the geodesic in two dimensions at the $step(n)$, for $n>0$, introduced in the above example in the $xy$-plane by (\ref{Geod}). Thanks to the geodesics between two antipodal points on spheres, we will rotate this geodesic around the $x$-axis between 0 and $2\pi$ to obtain all geodesics in three dimensions.
Using $N$ aligned balls (see Fig.2) and the geodesic coordinates given by (\ref{Geod}) we have:
\begin{equation}
\left\{
  \begin{array}{ll}
    x & =x \\
    y & = \varphi_{n}(x)\\
    z & =0
  \end{array}
\right.
\end{equation}
In general a rotation around the $x$-axis with angle $\theta$ is given by
\begin{equation}
\left\{
  \begin{array}{ll}
    x' &= x \\
    y' & = y\cos\theta-z\sin\theta\\
    z' & =y\sin\theta+z\cos\theta
  \end{array}
\right.
\end{equation}
By rotating the geodesic around the $x$-axis with angle $\theta\in[0,2\pi]$, we obtain
\begin{equation}\label{Ro}
\left\{
  \begin{array}{ll}
    x' &= x \\
    y' & = \varphi_{n}(x)\cos\theta\\
    z' & =\varphi_{n}(x)\sin\theta
  \end{array}
\right.
\end{equation}
 that provides all geodesics in three dimensions. Therefore for all $n>0$ the geodesics are sum of piecewise differentiable functions where the differentiable parts of the geodesics are great circle arcs, and these geodesics admit a countable number of  points of non differentiability at the $step(n)$.

The proof of iii) is obvious since $\cal M$ is an euclidian space at the $step(0)$.

\begin{cor}\label{C2}
In an expanding space-time  $\cal M$ that expands via discrete expansion of its basic elements, we have:

i) for $n>0$ there exists an infinity of geodesics between two distant points at the $step(n)$ and
all the geodesics vary together with the space-time expansion;

ii) for $n=0$ there exists a unique geodesic between two distant points.
\end{cor}

{\it Proof:} i) Since the geodesics are given by (\ref{Ro}) for all $\theta\in[0,2\pi]$, then for each value of $\theta$ we have a geodesic. Therefore there exists an infinity of geodesics between any two distant points in the expanding space-time $\cal M$ for $n>0$, and all these geodesics vary together with the space-time expansion (see Fig.3).

ii) For $n=0$ the conclusion is obtained by using i) of Theorem \ref{Chap2L2}.


\section{Expansion and Matter Distribution}

\subsection{Geodesics Location and Matter Distribution}

We know by Theorem \ref{Chap2L2} and Corollary \ref{cor4} that in the expanding space-time $\cal M$, the geodesics of light are located on the surface of the expanding basic elements as shown in Fig.\ref{Light2}. The Corollary \ref{C2} states that there exists an infinite number of geodesics that pass via expanding basic elements, and this is due to the existence of an infinite number of geodesics between two antipodal points on a given sphere. Since there is no privileged direction on the expansion of the basic elements of the expanding space-time $\cal M$ then the basic elements expand in all direction as balls, and since the light geodesics are located only on the surface of those expanding basic elements, then matter can be located on the surface of those expanding basic elements as illustrated in the left picture of Fig.\ref{Fig1}.

With this simulation we find out that matter is located on the surface of the considered spheres; and since matter in the physical world exists in 4 dimensional space-time, then the surface of the sphere in this simulation illustrates a distortion of 4 dimensional space-time. This can be understood from the local metric (\ref{M1}) of the considered expanding space-time. Indeed, the sequence $(a_i)$ of parameters of the metric (\ref{M1}) can be considered as new small dimensions (these dimensions are not perpendicular to the $x$, $y$, and $z$ axis, it can be deduced from the metric (\ref{M1})). The dimension of the simulated expanding space-time at the $step(n)$ is $4+n$ for all $n\geq0$, and the illustration of such a space-time in three dimensions necessitates the consideration of reduced dimensions.

\subsection{Invisibility of the Geometry of the Expanding Space-Time}

\begin{figure}[h!]
\begin{minipage}[t]{7cm}
\centering{\includegraphics[width=6cm]{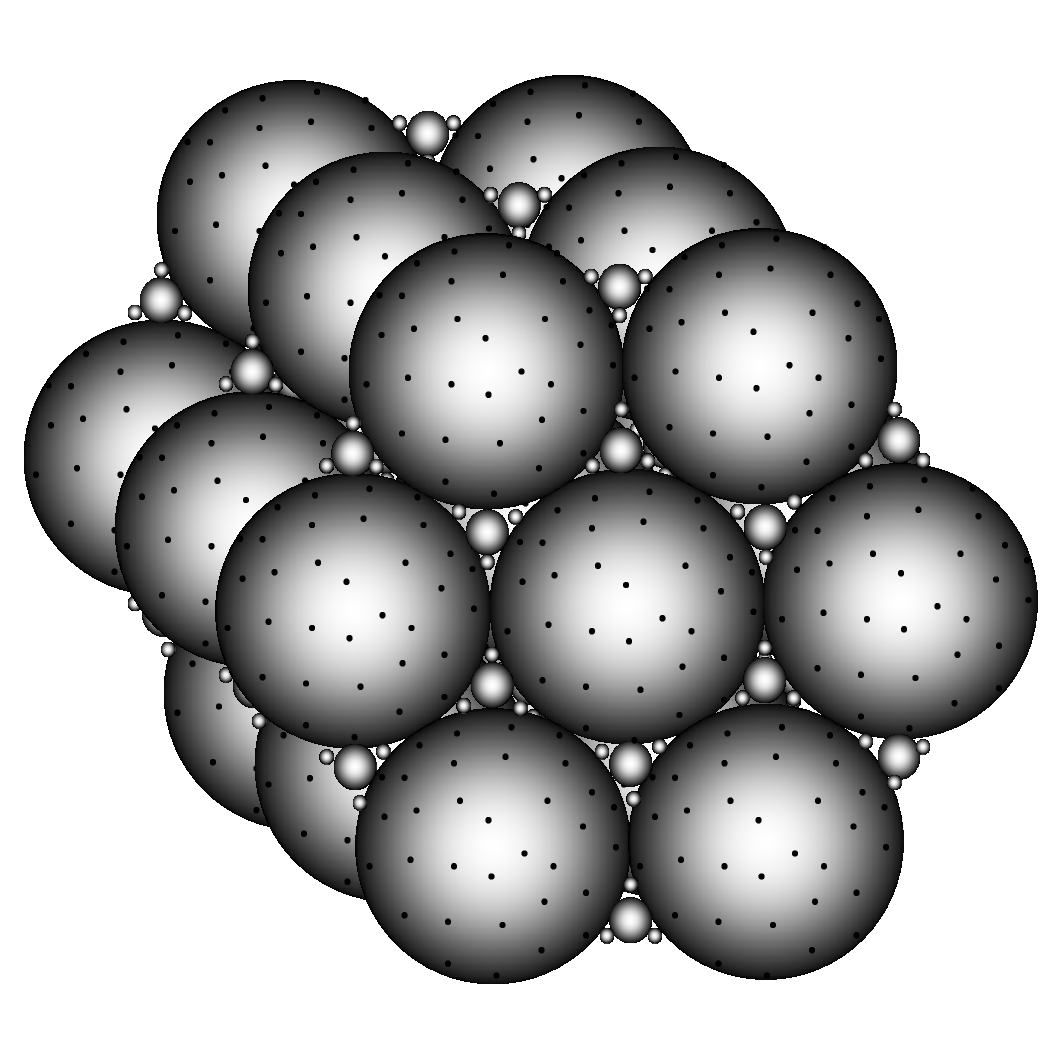}}
\vspace{.1mm}
\end{minipage}
\hspace*{\fill}
\begin{minipage}[t]{7cm}
\centering{\includegraphics[width=6cm]{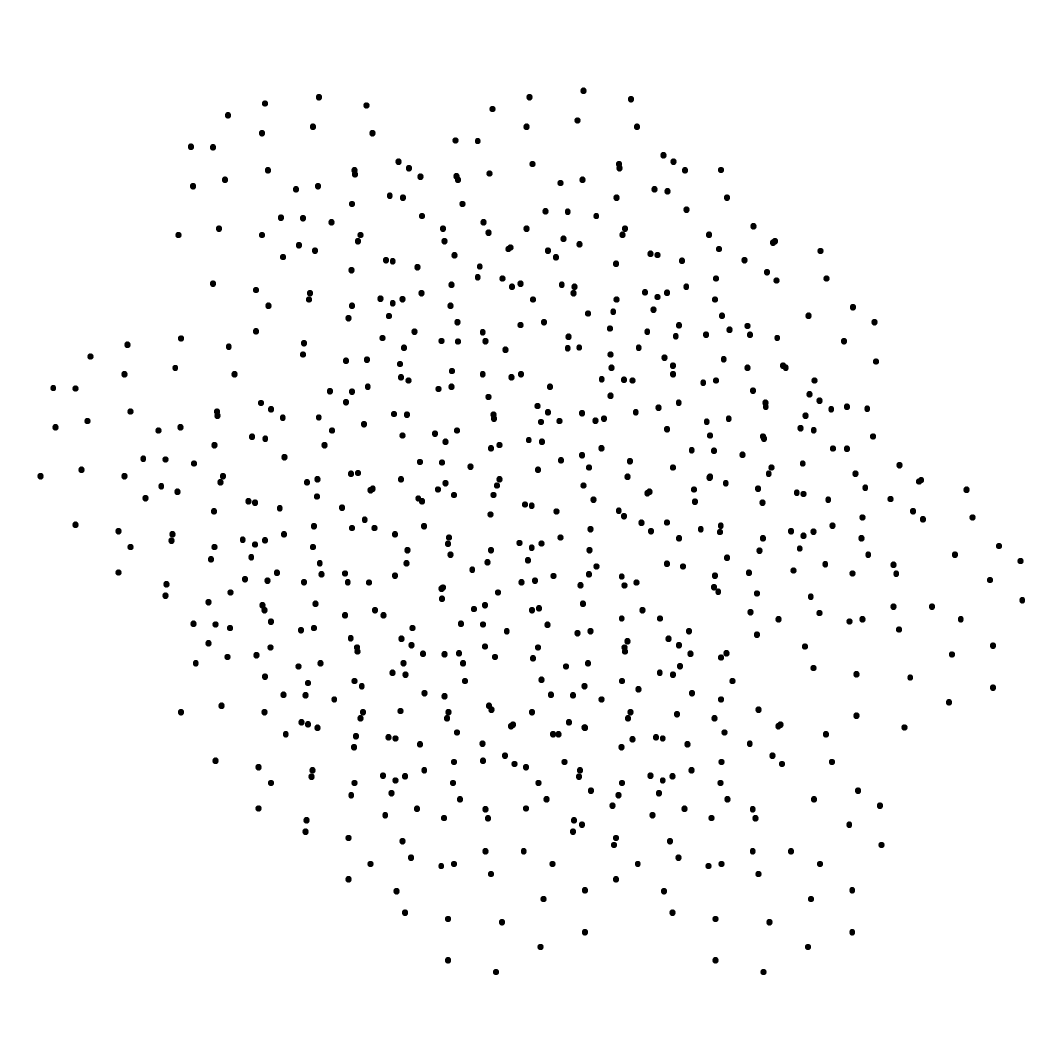}}
\vspace{.1mm}
\end{minipage}
\hspace*{\fill}
\caption{\footnotesize The left figure represents a portion of packed expanding basic elements at the present time (after 15 billion years from the Big Bang ($step(0)$)) where matter is illustrated as dots on their surface. Since the light geodesics are geodesic continuous contacts via those packed balls, then the right figure represents the same portion of packed expanding basic elements with matter and invisible geometry. Matter appears to be held in the expanding space-time with no visible pillars that one can see.}\label{Fig1}
\end{figure}

The geodesics of light are located on the surface of the expanding basic elements of the space-time $\cal M$. This means that the light will travel through expanding basic elements via geodesic continuous contacts that fluctuate on the boundary of the expanding basic elements, which makes the geometry of those expanding basic elements invisible since the basic elements neither reflect nor absorb light. They only bend light into a fluctuating motion, which makes the geometry completely invisible (see illustration in the right picture of Fig.\ref{Fig1}). The illusion to observe the information brought by light in a straight line was observed in cosmology for the first time in 1919 by Eddington and his team \cite{Chap3DED}. In general if the light is traveling in a given space-time following its curved geodesics, then the observer will have the illusion to interpret the location of the information brought by light in a straight line position (the location of a given galaxy for example). That is why in the expanding space-time $\cal M$ the curved geodesics of light cannot be observed, which makes the geometry of $\cal M$ invisible, and for any observer located in such expanding space-time, matter appears to be held with no visible pillars that one can observe (similar interpretation can be found in \cite{Chap3BF2}).

\begin{figure}[h!]
\begin{minipage}[t]{7cm}
\centering
\includegraphics[width=8cm]{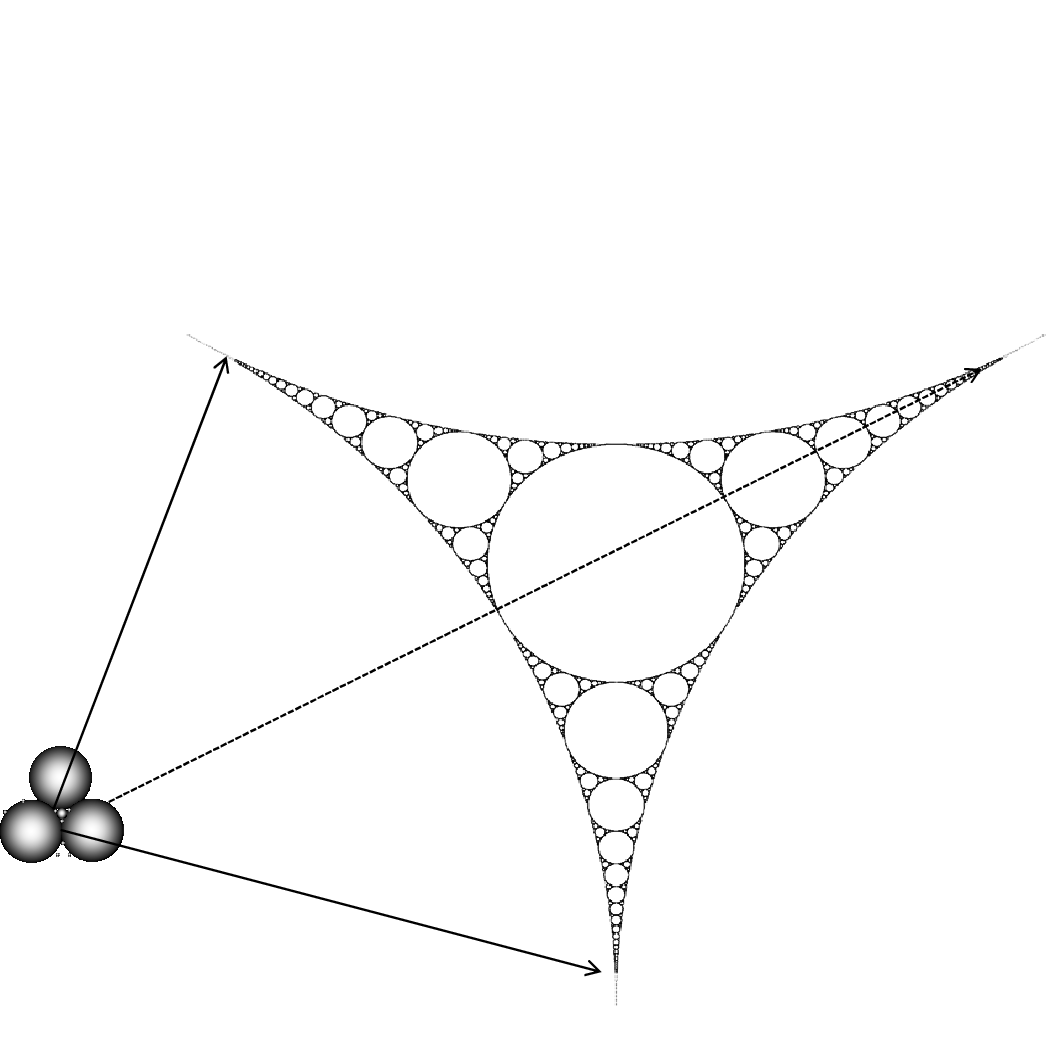}
\vspace{.1mm}
\end{minipage}
\hspace*{\fill}
\begin{minipage}[t]{7cm}
\centering
\includegraphics[width=7cm]{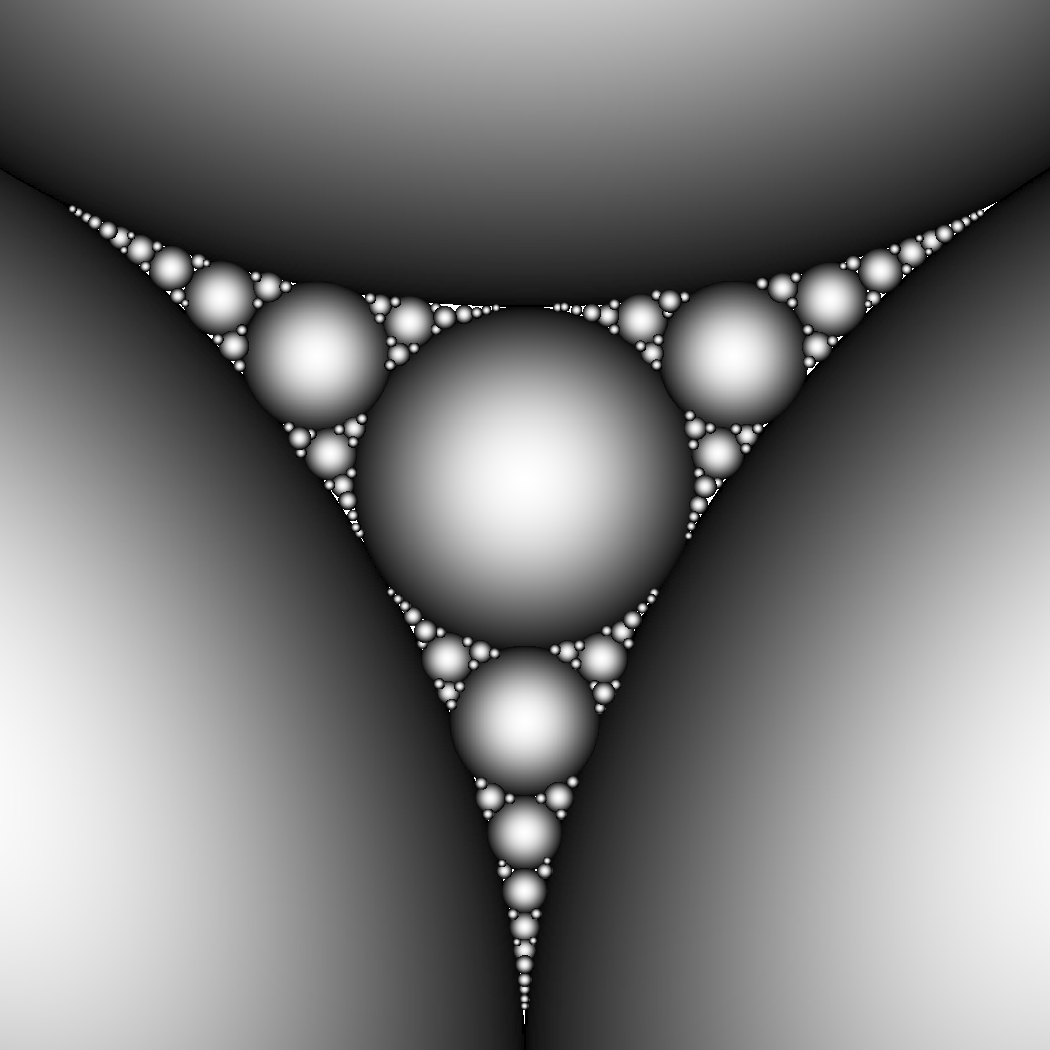}
\vspace{.1mm}
\end{minipage}
\hspace*{\fill}
\caption{\footnotesize The appearance of an Apollonian gasket between each three packed expanding basic elements gives the fractal character to the shape of the expanding space-time $\cal M$. Each Apollonian gasket is constituted with packed expanding balls with different sizes as illustrated here.}\label{Light6}
\end{figure}

\begin{rem}
1) If three packed basic elements of the expanding space-time $\cal M$ expand, the density of the primordial space-time means that there is no holes between them, then other new balls with different sizes will appear to fill up the interstices. This induces the appearance of an Apollonian gasket between each three expanding basic elements that expand as illustrated in Fig.\ref{Light6} and then we have appearance of new structures at different steps from the space-time expansion, which confirms the fractal character of the space-time geometry.

2) The appearance of an Apollonian gasket between each three packed expanding basic elements creates locally a non homogeneity of the space-time, meanwhile the simultaneous expansion of the primordial basic elements (at the $step(0)$) of the space-time creates the homogeneity of the space-time at large scale.

3) The structure of space-time shape described by the expanding space-time $\cal M$ looks like the atomic structure of matter, and it seems not strange to find a similarity on the way they are packed to be stable. Matter is constituted by packed atoms, meanwhile in the expanding space-time $\cal M$, matter is held by packed expanding balls.
\end{rem}

\subsection{Curvature and Topology of the Expanding Space-Time}

The picture of the expanding space-time $\cal M$ conveys a very simple idea: the shape of the space-time is obtained via packed expanding balls with different sizes. The expansion of those packed balls creates the universe expansion, where locally the curvature can be obtained straight forward via the curvature of the expanding spheres of different sizes. Indeed, it is known that the normal curvature of a sphere of radius $R$ is everywhere and in all direction given by the constant $R^{-1}$ so that the Gaussian curvature is given by  $1/R^2$, meanwhile the mean curvature is given by $2/R$ (see\cite{Chap3DFN} p.89). The curvatures (normal curvature, Gaussian curvature, and the mean curvature) of the sphere are functions of its radius, and if the radius varies the curvatures will also vary. This leads to the conclusion that the expanding space-time $\cal M$ has locally and globally a variable curvature. Globally at large scale (if we consider the radius of the primordial expanding basic elements at present time) the curvature will be approximatively close to zero since the primordial balls were expanding for 15 billion years (that is to say the radius of the primordial expanding basic elements is considered to be very large today), which indicates that the space-time is almost flat, meanwhile if we consider the measure of radius of balls from the Apollonian gasket that exists between each packed three primordial basic elements, we will find a variable curvature less than 1.\pesp

This simulation of the expansion of the space-time via expansion of its basic elements takes into account the density of the space-time and leads to the appearance of other new basic elements, with different sizes in any interstice, that form Apollonian gaskets of packed expanding balls (see Fig.\ref{Light6}).
The appearance of Apollonian gasket between any three packed basic elements creates the fractal character of the space-time and simulates locally the non-homogeneity of the expanding space-time; meanwhile the simultaneous expansion of the primordial basic elements simulates the homogeneity and isotropy of the expanding space-time at large scale, which is consistent with the cosmological principles \cite{Chap3PEE}.\pesp

 The location of matter on the surface\footnote{this surface is an illustration of a distorted space-time of dimension 4 since the dimension of the whole space (space-time including holes) is $n+4$ at the $Step(n)$ for all $n\geq0$.} of the expanding balls means that the open balls are empty since they are a forbidden region for matter, they correspond to holes in the space-time. Therefore the appearance of Apollonian gasket between any three packed basic elements creates holes and then makes the topology variable. This simulation presents a concrete example of a discrete space-time expansion with variable topology, and can be used in fractal topology (\cite{Chap3PO},\cite{Chap3PO2}) as a case study.

\section{Infinity of Geodesics and Application in Physics}

It is known that wave optics predicts phenomena such as polarization, interference and diffraction, which are not explained by geometric optics. Quantum mechanics prohibits localization in the quantum world based on the Heisenberg uncertainty principle that makes simultaneous measurements of position and momentum of a quantum physical system impossible to obtain with accuracy. The use of an infinity of transverse fluctuating geodesics found in this simulation (see Fig.\ref{Fig.001}) is compatible with the non localization in quantum mechanics. Indeed, the fluctuating geodesics (Corollary \ref{C2}) are sum of piecewise differentiable functions with a countable number of points of non differentiability: the antipodal points. The parts that are represented by differentiable functions between any two antipodal points are great circle arcs, and their number is infinite; even if one can determine the velocity given by the derivative of the  differentiable functions, it is impossible to determine which path is used among the infinite number of geodesics between two antipodal points. Moreover the antipodal points are points of non differentiability that give the location  while it is impossible to determine the velocity at these points. Then it is impossible to know simultaneously with accuracy the location and the momentum of a physical system that follows these geodesics in its motion.

A prediction of light diffraction using these infinity of fluctuating geodesics was recently obtained (see \cite{FB25}), and a prediction of interference pattern was obtained with these infinity of fluctuating geodesics using a highly reproducible superimposition of geodesics graphs (see \cite{FB35}), which is an asset for geometric optics to provide interpretation for physical phenomena that thought to be impossible to explain.

A prediction about polarization in geometric optics can be obtained using the infinity of fluctuating geodesics:

\subsection{Infinity of Fluctuating Geodesics and the Ray Approximation}

\begin{figure}[h]
\centering
\includegraphics[width=9cm]{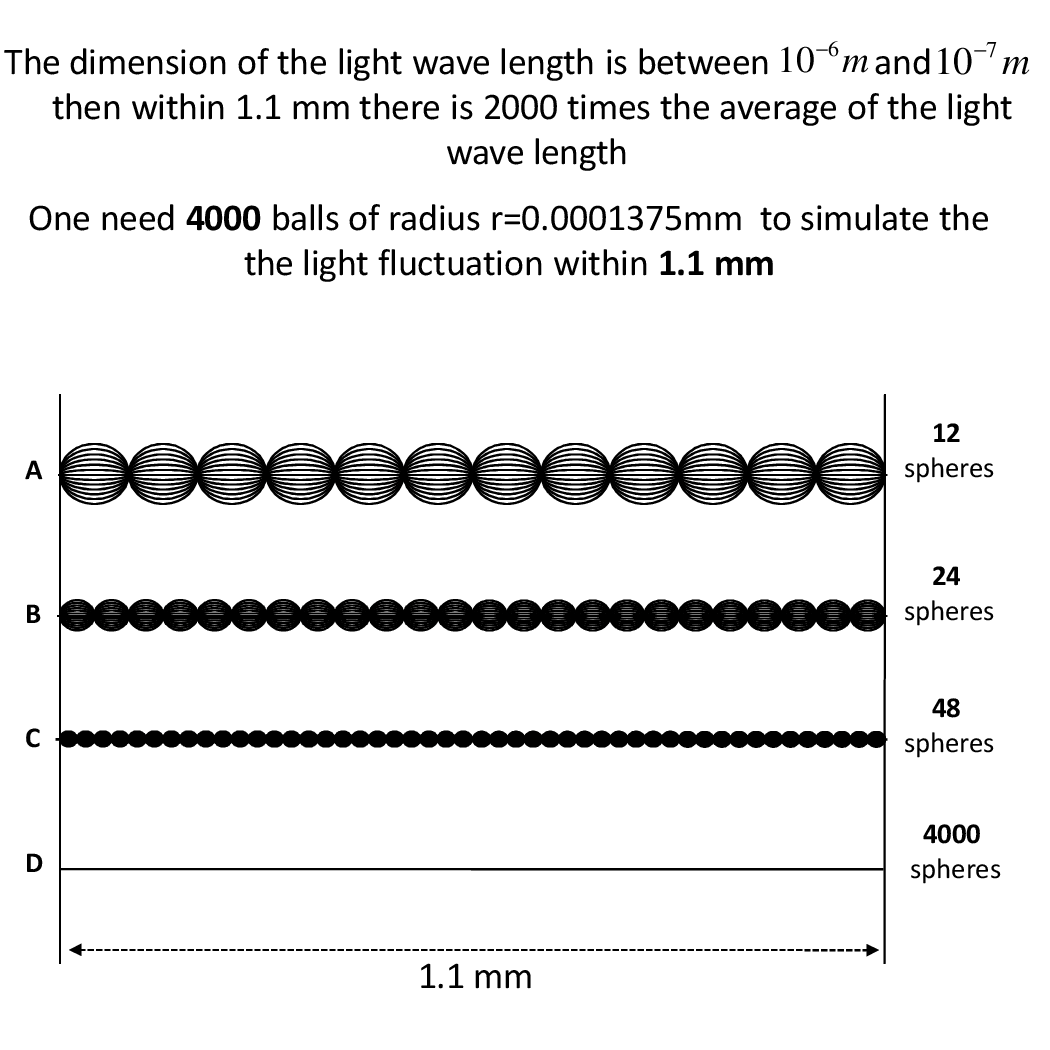}
\caption{\footnotesize  Within $1.1 mm$, one needs to accumulate in straight line: A) 12 spheres of radius $r_A\simeq0,0458333\ mm$, B) 24 spheres of radius $r_B\simeq0,0229166\ mm$, C) 48 spheres of radius $r_C\simeq0,0114583\ mm$, D) 4000 spheres of radius $r_D\simeq0,0001375\ mm$. The 4000 spheres of radius $r_D$ appear as a straight line, the more the number of accumulated spheres within $1,1 mm$ increases, the smaller the dimension of the spheres is.}\label{Fig.00}
\end{figure}

The infinity of fluctuating geodesics can explain why the laser beam manifests an observed straight line geodesic. Indeed, in Corollary \ref{C2} we proved that there exists an infinity of fluctuating geodesics between two distant locations defined by (\ref{Ro}), and the use of these geodesics at the average scale of the light wave length (for $10^{-7}m\leq 4r\leq 10^{-6}m$ in (\ref{Ro})) makes the fluctuation of the geodesics undiscernible at the macroscopic scale without aid of magnifying devices (see Fig.\ref{Fig.00}). The small size of the geodesics fluctuation, with respect to the length scale on which objects are sufficient big enough to be visible without the aid of magnifying devices, makes the infinity of fluctuating geodesics perceived as a straight line, meanwhile a microscopic view of them reveals an infinity of transverse fluctuating geodesics.

Performing optics research that includes the study of how light interacts with specific materials using the model of infinity of transverse fluctuating geodesics may lead to perform research in fiber-optic cables for example. Fiber-optic cables carry information between two places using optical technology (light-based).

The infinity of transverse fluctuating light geodesics at the quantum scale could be used to improve the quality of transmission in fiber-optics, the more the boundary is smooth at the quantum scale, the better the transmission is. Better improvement might be expected when regularity reaches the small scale level using the nanotechnology, and the infinity of fluctuating geodesics of light at the quantum scale might explain it. The light would be considered entirely transmitted using the infinity of fluctuating geodesics if the beam of light is entirely reflected using all the infinity of geodesics, if some of them are reflected and others are obstructed the transmission is not complete, which affects the quality of transmission.

\subsection{Prediction of Polarization in Geometric Optics}

The use in physics of the ray approximation of light (when we neglect the wave nature of light) precises the direction of the light propagation. However, some phenomena such as polarization, interference and diffraction (the study of these phenomena constitutes the wave optics) reveal the wave characteristics of light and the limit of the geometric optics.

\begin{figure}[h]
\centering
\includegraphics[width=7cm]{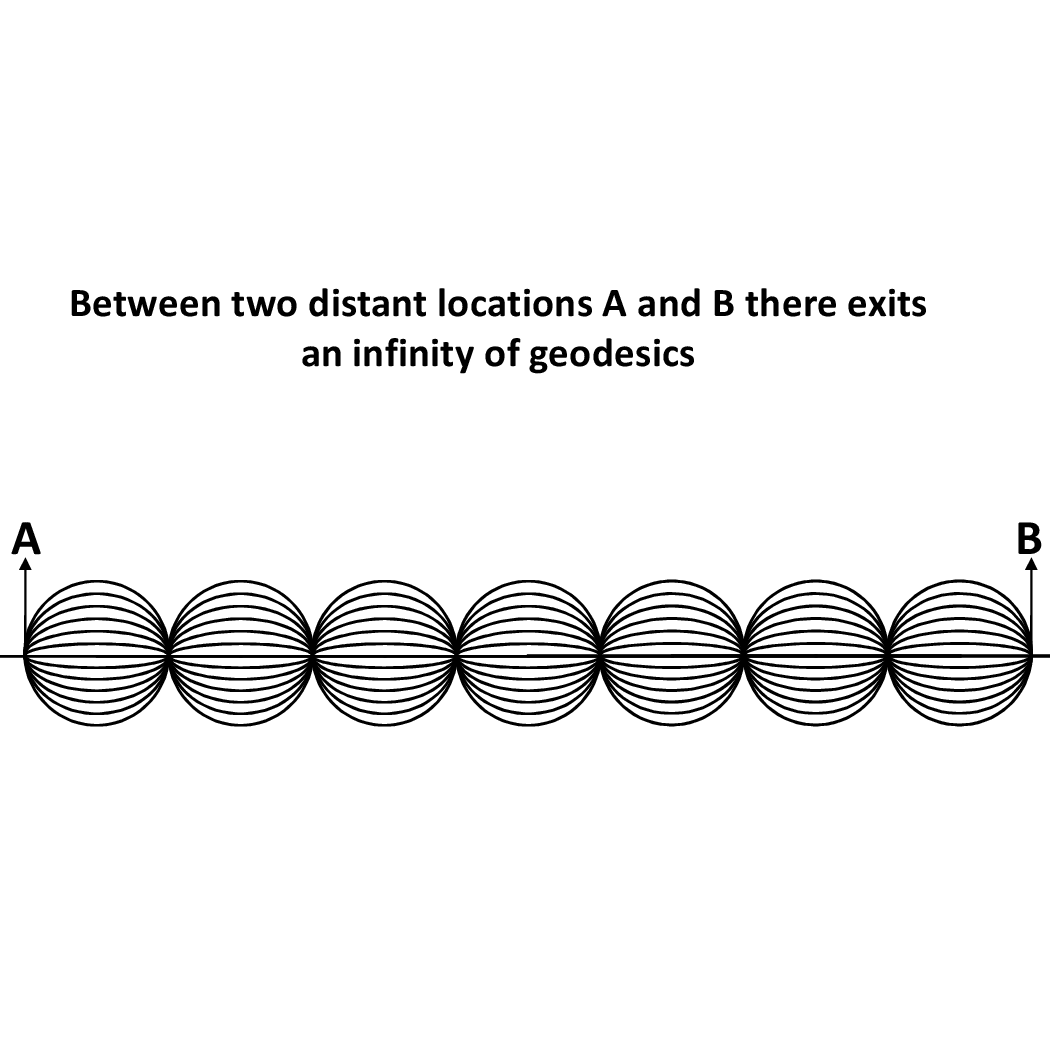}
\caption{\footnotesize  Existence of infinity of transverse fluctuating geodesics between two distant locations A and B in an expanding homogeneous and isotropic space-time.}\label{Fig.001}
\end{figure}

The simulation within this paper reveals that the path of least time for the light in an expanding homogeneous and isotropic space-time cannot be a straight line, it is rather an infinity of transverse fluctuating geodesics (Corollary \ref{C2}), see illustration in Fig.\ref{Fig.001}.

In general, light splits into two beams that travel in different directions at different speeds (called double refraction) in crystalline materials. Calcite shows a double refraction: if one observes calcite through a polarizing filter, the doubled images blink in and out as the polarizer is rotated, which confirms that the two images are made of light polarized in perpendicular directions \cite{HRW5}.

Polarization was thought as a general property of waves that describes the orientation of their oscillations. However with the model of the infinity of transverse fluctuating geodesics defined by (\ref{Ro}) and represented in Fig.\ref{Fig.001}, polarization of light that travels between two distant locations using the infinity of fluctuating geodesics can be explained to describe the orientation of the transverse fluctuation that can be followed by the light. It describes the orientation of the fluctuation in the plane perpendicular to the light direction between two distant positions. The fluctuation may be oriented in a single direction (linear polarization), or the fluctuation direction may rotate in the motion of the physical system (circular or elliptical polarization) similar to the wave polarization model. The fluctuation of the geodesics can be limited to a single plane called the plane of fluctuation if we obstruct the other planes of fluctuation.

\subsection{Prediction of acceleration in the cosmic inflation model}

Taking into account the fundamental principles of cosmology, we have built a simulation of an homogeneous and isotropic expanding space-time via discrete expansion of its basic elements to figure out the consequence of this expansion on the geodesics.
The properties of this simulation can be used in cosmology to explain the earlier acceleration of the universe expansion in the cosmic inflation model, and proves that any consideration of homogeneity and isotropy in the simulation of the space-time expansion, via expansion of its basic elements, leads to the existence of an earlier acceleration (inflation) at the beginning of space-time expansion.\pesp

The expansion of the universe means that the proper distance between a pair of well separated galaxies is increasing with time, that is the galaxies are receding from each
other. This observed expansion means that all matter were very close to each other some 15 billion years ago, and matter were assembled in a very hot and extremely dense region in an absolute limit on size and density allowed by quantum physics. Under these conditions, the first question is how anything that dense could ever expand since it is known that an extremely strong gravitational field must exist in such conditions, and it could turn the extremely dense compact region into a black hole that could prevent matter recession from each other, and turn everything into a singularity, which would sharply stop the process of the universe expansion.
A short period of extremely rapid expansion is therefore needed to avoid this scenario and that was the suggestion of the cosmic inflation model. Cosmologists introduced this idea of a short and earlier acceleration to address some problems that cannot be addressed within the context of the standard Big Bang

The inflationary hypothesis was developed by the physicists A. Guth and A. Linde in 1980 to solve several important problems in cosmology such as horizon problem and flatness problem (\cite{Gut},\cite{GutYoug},\cite{lind}). It provides a vision of the distribution of matter at large scale structure, quantum fluctuation and interpretation of the cosmic microwave background radiation (\cite{PEI5},\cite{SP5}), the non existence of magnetic monopole, and others (\cite{kinn},\cite{Chap3PEE}).

A short acceleration\footnote{An exponential expansion of the space-time was found by Willem de Sitter \cite{WDS5} in 1917 using the general relativity to construct a cosmological model. For more than half a century, the De Sitter model was considered as a mathematical curiosity with no relevance to the real Universe. Nowadays it is considered as a cornerstone of inflationary cosmology.} during a fraction of second after the Big Bang is needed to insure the universe expansion and address some observed problems. Nevertheless, the cosmic inflation model postulates this acceleration despite the fact that the mechanism that creates this inflation is not known yet.

The simulation of the expansion of an homogenous, isotropic and continuous space-time via discrete expansion of its basic elements provides a possible explanation of the mechanism of such short inflation at a very early epoch of the universe existence.

Indeed, an Euclidian space-time with zero curvature set of accumulated points without depth (their dense accumulation defines the space) is taken as a starting stage of the modeled space-time (initial condition). The simulation of the space-time expansion is generated via the expansion of its basic elements: the Euclidian space-time expands via simultaneous expansion of its basic elements, that is to say points increase simultaneously their depth in all directions as balls. The result is an infinite family of packed expanding balls that simulates the expanding space. This expansion is simulated via the quantification of a discrete local expansion from the $Step(0)$ (the initial space) to the $Step(n)$ for all $n>0$.\pesp

Taking into account the density of the space-time, other new basic elements with different sizes expand in any interstice, which creates an Apollonian gasket of packed expanding balls with different sizes. The appearance of Apollonian gasket between any three packed basic elements creates the fractal character of the space-time and simulates locally the non-homogeneity of the expanding space-time; meanwhile the simultaneous expansion of the primordial basic elements simulates the homogeneity and isotropy of the expanding space-time at large scale, which is consistent with the cosmological principles.\pesp

Based on the definition of the discrete expansion given in Definition \ref{S0}, the mechanism of the discrete expansion of the primordial basic elements can be summarized step by step as follows:

\begin{itemize}
  \item from the step $Step(0)$ to the $Step(1)$, the simultaneous expansion transforms the basic elements of the Euclidian space-time from dimensionless points to tiny open balls of radius $r_1=a_1$ (balls of dimension 3).
  \item from the $Step(1)$ to the $Step(2)$, the simultaneous expansion transforms open balls of radius $r_1=a_1$ to open balls of radius $r_2=a_1a_2$.
  \item from the $Step(n)$ to the $Step(n+1)$ for all $n \geq1$, the simultaneous expansion transforms open balls of radius $r_{n}=\di\prod_{i=1}^{n}a_i$ to open balls of radius $r_{n+1}=\di\prod_{i=1}^{n+1}a_i$ (bigger balls).
\end{itemize}

 If we consider the parameter $a_i$ as a new dimension\footnote{From the metric (\ref{M1}) and the Definition \ref{S0} the discrete expansion, one can understand that these new dimensions are small and not perpendicular to those of $\rR^3$, and that the discrete expansion is an expansion in dimensions.}, then this discrete simulation of expansion transforms:
\begin{itemize}
  \item dimensionless points into  open balls of dimension 3 from $Step(0)$ to $Step(1)$
  \item open balls of dimension 3 into open balls of dimension 3+1 from $Step(1)$ to $Step(2)$
  \item open balls of dimension $3+(n-1)$ into open balls of dimension $3+n$ from $Step(n)$ to $Step(n+1)$ for all $n \geq1$.
\end{itemize}

The discrete simultaneous expansion can be seen as an expansion in dimensions, where the expansion in dimensions from the $Step(0)$ to the $Step(1)$ is different from the expansion in dimensions from $Step(n)$ to $Step(n+1)$ for all $n \geq1$. This difference is the source of the acceleration in the space expansion at the beginning of the space-time expansion. The distance between two distant basic elements becomes infinite at the $Step(1)$, which prevents the extremely strong gravitational field to turn the extremely dense compact region into a black hole that prevents expansion. In this simulation the deformation of the space-time from the $Step(0)$ to the $Step(1)$ corresponds to a short period of extremely rapid expansion (see Remark \ref{Rem2}, i)), this is due to the variation of big  dimensions from 0 to 3 (that correspond to $x$, $y$ and $z$), meanwhile the variation from the $Step(n)$ to $Step(n+1)$ for all $n \geq1$ is made via  variation of the small dimensions that corresponds to the new small parameters $a_i$ for $i\geq1$ which  are not perpendicular to the  $x$, $y$, and $z$ axis.

\subsection{Conclusion}

We have conducted a simulation of an expanding space-time that expands through expansion of its basic elements, that leads to obtain an expanding space-time consistent with the cosmological principles. The mechanism of this expansion was chosen by discrete transformation of the basic elements from points to expanding open balls to identify the characteristics and  effects of this local transformation on space-time geodesics. We find out the following:

\begin{itemize}
  \item the straight line geodesics cannot exist in such an expanding space time;
  \item the geodesics are bend by the basic elements expansion to form fluctuating geodesics;
  \item there exists an infinity of geodesics between two distant locations;
  \item as a consequence of the geodesics fluctuation, matter exists only on the surface of the expanding open balls of dimension $n+4$, which suggests that the interior of those balls is forbidden for matter, and where the surface illustrates a distortion of 4 dimensional space time;
  \item the expansion of the basic elements creates the variation of topology;
  \item the geometry of the expanding space-time is variable with expansion;
  \item the basic elements of the expanding space-time neither reflect nor absorb light, they only bend light into a fluctuating motion, which makes the geometry of the homogeneous and isotropic expanding space-time completely invisible.
\end{itemize}

In \cite{Chap3PO},\cite{Chap3PO2} the author proved that the variation of topology creates the space-time expansion, and within this simulation we have proved that the expansion of the space-time creates the variation of topology. This confirms the existence of a causality between space-time expansion and  variation of the space-time topology and suggests that more implication of topology in cosmology is needed for better understanding.
This simulation with basic conditions (homogeneity, isotropy and expansion) allows to understand the expansion of a space-time generated by the expansion of its basic elements, and leads to figure out the causal relation between expansion and topology. The non existence of straight line geodesics may have an impact on the interpretation of different phenomena in physics (such as prediction of diffraction, interference and polarization in geometric optics). The determination of the geodesics in such an expanding space time provides an optimal location of matter distribution at large scale as well as the invisibility of its geometry. The existence of an infinity of geodesics between any two distant locations in a fractal space-time was first predicted by Nottale in \cite{Chap3NT2}. The notion of fractal space-time in Nottale's theory of scale relativity (\cite{Chap3NT1},\cite{Chap3NT2},\cite{Chap3NT3}) was used without mathematical characterization of its geometry and its topology. The dynamic of the space-time defined by an accumulation of infinity of balls can be studied using this simulation. This case study is the basis for other simulations with more conditions consistent with observations that could be elaborated for a better understanding of the universe dynamics.

\bibliography{apssamp}

\begin{thebibliography}{99}


\bibitem{Chap3BF1} F.Ben Adda, \textit{Mathematical Model for Fractal Manifold}, International Journal of Pure and Applied Mathematics, Volume 38, {\bf N2} (2007), p.159-190.
\bibitem{Chap3BF2} F.Ben Adda,  \textit{New Understanding of the Dark Energy in the Light of New Space-Time}, Invisible Universe: Proceeding of the Conference, AIP Conference Proceedings, \textbf{1241} (2010), 487-496.
%
\bibitem{FB25} F.Ben Adda, \textit{A New Fundamental Factor in the Interpretation of Young's Double-Slit Experiment}, Class-Ph/viXra:1406.0015.
%
\bibitem{FB35} F.Ben Adda, \textit{Paths of least time for quantum scale and geometrical interpretation of light diffraction}, Gen-Ph/arXiv:1409.4257v2.
%
\bibitem{Chap3BL} H.F.Blichfeldt, \textit{The minimum value of quadratic forms and the closest packing of spheres}, Mathematische Annallen, {\bf 101}, N 1 (1929), 605-608.
%
\bibitem{Chap3BOR} K.B\"or\"oczky, \textit{Packing of spheres in spaces of constant curvature}, Acta Mathematica Hungarica, {\bf 32}, N 3-4 (1978), 243-261.
%
\bibitem{Chap3BUR} J.A.C.Burlack, R.A.Rankin, A.P.Robertson: \textit{The packing of spheres in the spaces lp},
Proc. Glasgow Math. Assoc. {\bf 4} (1958), 22-25.
\bibitem{Chap3BCNN} J.Banks, J.Carson, B.Nelson, D.Nicol. \textit{Discrete-Event System Simulation}. (Prentice Hall 2001), p. 3, ISBN 0-13-088702-1.
%
\bibitem{WDS5} W.De Sitter, \textit{ On Einstein's Theory of Gravitation and Its Astronomical Consequences: Third Paper}. Monthly Notices of the Royal Astronomical Society, {\bf 78} (1917), 3–28.
%
\bibitem{Chap3DFN} A.B.Dubrovin, A.T.Fomenko, S.P.Novikov,  \textit{Modern Geometry, Methods and Applications}, Springer-Verlag, (New York Inc. 1992).
%
\bibitem{Chap3DED} F.W.Dyson, A.S.Eddington, C.R.Davidson, \textit{A Determination of the Deflection of Light by the Sun's Gravitational Field, from Observations Made at the Solar eclipse of May 29, 1919}, Phil. Trans. Roy. Soc. A {\bf 220} (1920), 291-333.
%
\bibitem{Gut} A.Guth, \textit{Inflationary universe: A possible solution to the horizon and flatness problems} Phys. Rev. D {\bf 23} (1981), 347.
%
\bibitem{GutYoug} A.H.Guth and S.Y.Pi, \textit{Fluctuations in the New Inflationary Universe}, Phys. Rev. Lett. {\bf 49} (1982), 1110.
%
\bibitem{Chap3GA} C.F.Gauss, \textit{Besprechung des buchs von L.A.Seeber: Untersuchungen \"uber die eigenschaften der positiven tern\"aren quadratischen formen usw}, G\"ottingsche Gelehrte Anzeigen, {\bf 2} (1876), 188-196.
%
\bibitem{Chap3HUB1} E.Hubble, \textit{NGC 6822, a remote stellar system}, Astrophysics Journal {\bf 62} (1925), 409-433.
%
\bibitem{Chap3HUB2} E.Hubble, \textit{Extragalactic nebulae}, Astrophysical Journal {\bf 64} (1926), 321-369.
%
\bibitem{Chap3HUB3} E.Hubble, \textit{A relation between distance and radial velocity among extra-galactic nebulae}, PNAS {\bf 15} (3) (1929), 168-173.
%
\bibitem{HRW5} D.Halliday, R.Resnick, J.Walker, {\em Fundamental of physics}, John Wiley and sons, Inc (2008), ISBN 978-0-471-75801-3.
%
\bibitem{kinn} W.H.Kinney , \textit{Cosmology, Inflation and the Physics of Nothing}, Techniques and Concepts of High-Energy Physics XII, NATO Science Series Volume {\bf 123} (2003),  189-243.
%
\bibitem{Chap3LA} J.L.Lagrange,\textit{Recherches d'arithm\'{e}tique}, Nouv. Mem. Acad. Roy. Sc. Belles Lettres,
    (Berlin 1773), 265-312, (Oeuvres compl\`{e}tes Vol III, 693-758).
%
\bibitem{Chap3GL} G.Lema\^ itre, \textit{Un univers homog\`{e}ne de masse constante et de rayon croissant rendant compte de la vitesse radiale des n\'{e}buleuses extra-galactiques}, Annales de la Soci\'{e}t\'{e} Scientifique de Bruxelles, A{\bf 47} (1927), 49-59.
%
\bibitem{lind} A.D.Linde , \textit{A new inflationary universe scenario: A possible solution of the horizon, flatness, homogeneity, isotropy and primordial monopole problems}, Physics Letters B, {\bf 108}, Issue 6 (1982), Pages 389-393.
%
\bibitem{Chap3LI} J.H.Lindsey, \textit{Sphere packing in $\rR^3$}, Mathematica, {\bf 33} (1986), 137-147.
%
\bibitem{Chap3MIL} J.Milnor, \textit{Hilbert' problem 18: On the crystallographic groups fundamental domains and on sphere packings}, Proc. of Symp. in Pure Math. AMS {\bf 28} (1976), 491-506.
%
\bibitem{Chap3MO} H.Morall, J. Phys. A: Math. Gen. {\bf 27} (1994), 7785-7791.
%
\bibitem{Chap3NT1} L.Nottale, \textit{The Theory of Scale Relativity}, Int. J. Mod. Phys. A {\bf 7} (1992), 4899-4936.
%
\bibitem{Chap3NT2} L.Nottale, \textit{Fractals and the quantum theory of space-time}, Int. J. Mod. Phys. A {\bf 4} (1989), 5047-5117.
%
\bibitem{Chap3NT3} L.Nottale, \textit{Fractal space time and microphysics}, World Scientific (1993).
%
\bibitem{Chap3PEE} P.J.E.Peebles, \textit{Principles of physical cosmology}, Princeton Series in Physics, (Princeton University Press 1993), ISBN 0-691-07428-3.
%
\bibitem{PEI5} H.V.Peiris, et al., \textit{First Year Wilkinson Microwave Anisotropy Probe (WMAP) Observations:  Implications for Inflation}, ApJS, {\bf 148} (2003), 213.
%
\bibitem{Chap3PO} H.Porchon, \textit{Fractal Topology Foundations}, Topology and its Applications, {\bf 159}, 14 (2012), 3156-3170.
\bibitem{Chap3PO2} H.Porchon, \textit{Expanding Topological Space, Study and Applications}, Topology and its Applications, {\bf 160}, 16 (2013), 2121-2139.
%
\bibitem{Chap3RA} R.A.Rankin, \textit{On the Closest Packing of Spheres in n Dimensions}, The Annals of Mathematics, Second Series, {\bf 48}, No. 4 (1947), 1062-1081.
%
\bibitem{Chap3RO} C.A.Rogers, \textit{The packing equal spheres}, Proc. London Math. Soc., V S3-8 {\bf 4} (1958), 609-620.
%
\bibitem{Chap3SC} G.D.Scott, \textit{Packing of spheres: Packing of equal spheres}. Nature, {\bf 188} (1960), 908-909.
\bibitem{Chap3SO} F.Soddy, Nature, 139 (1937), 77-79.
\bibitem{Chap3SB} J.A.Sokolowski, C.M.Banks, \textit{Principles of Modeling and Simulation}. Hoboken, NJ: Wiley (2009). p. 6. ISBN 978-0-470-28943-3.
%
\bibitem{SP5} D.N.Spergel et al.,  \textit{Three-year Wilkinson Microwave Anisotropy Probe (WMAP) observations: Implications for cosmology}, ApJS, {\bf 170} (2007), 377.
%

\end{thebibliography}

\end{document}